\documentclass[acus]{epac98}

\usepackage{epsfig}

\newcommand{\bit}{\begin{Itemize}}
\newcommand{\eit}{\end{Itemize}}

\begin{document}
\title{WAKE FIELD EFFECT ANALYSIS IN APT LINAC}

\author{Sergey S. Kurennoy \\
LANSCE-1, Los Alamos National Laboratory,
Los Alamos, NM 87545, USA}

\maketitle

\begin{abstract} 
The 1.7-GeV 100-mA CW proton linac is now under design for the 
Accelerator Production of Tritium (APT) Project \cite{APT}. The APT 
linac comprises both the normal conducting (below 211 MeV) and 
superconducting (SC) sections. The high current leads to stringent 
restrictions on allowable beam losses ($<$ 1~nA/m), that requires 
analyzing carefully all possible loss sources. While wake-field 
effects are usually considered negligible in proton linacs, we 
study these effects for the APT to exclude potential 
problems at such a high current. Loss factors and resonance 
frequency spectra of various discontinuities of the vacuum 
chamber are investigated, both analytically and using 2-D and 3-D 
simulation codes with a single bunch as well as with many bunches. 
Our main conclusion is that the only noticeable effect is the HOM 
heating of the 5-cell SC cavities. It, however, has an acceptable 
level and, in addition, will be taken care of by HOM couplers.
\end{abstract}

\section{Introduction}

A wake field analysis for a high-intensity accelerator typically
includes wake and/or coupling impedance computations, and following
calculations of loss factors and heating due to various elements of 
the vacuum chamber, as well as a study of possible instabilities.
Beam coupling impedances and loss factors can be obtained from wake 
fields computed by time-domain codes like ABCI \cite{ABCI} and 
MAFIA \cite{MAFIA}. However, this approach works only for an 
ultrarelativistic bunch, with $\beta=v/c=1$, because of difficulties
formulating open boundary conditions for $\beta < 1$ in time domain. 

There are two specific features of the wake-field analysis in proton 
(or H${}^-$) high-intensity linacs. First, $\beta$ is significantly 
less than 1 for the most part of the machine. Results at $\beta=1$, 
while provide useful estimates, can be quite different from those at
the design $\beta$ values in some cases, see, e.g., \cite{SK}. 
In particular, the resonance impedances and corresponding loss factors 
can strongly depend on $\beta$. 
Frequency-domain calculations provide an accurate answer for a given 
$\beta < 1$, but typically they are limited to just a few lowest modes. 
Second, the beam in high-intensity linacs is either CW, or consists of
macropulses containing many regularly spaced bunches. As a result,
the beam frequency spectrum is concentrated only near the multiples 
of the bunch repetition frequency $f_b$. Of course, the spectrum 
envelope is defined by the bunch shape, but due to short bunch length 
it rolls off at frequencies many times higher than $f_b$.    

Therefore, an important question to consider is whether any higher 
mode has its frequency close to a multiple of $f_b$. The presence of
such modes, especially at relatively low frequencies, can lead to
undesired coherent effects. We use time-domain computations with 
multiple bunches to answer this question. The idea is to apply
a standard time-domain code with a few identical bunches at $\beta=1$, 
but to set the bunch spacing $s$ to $s=c/f_b$ for having the correct
bunch repetition frequency. Since the resonance frequencies are 
essentially independent of $\beta$, so is a conclusion from such 
simulations. In this note, we concentrate only on this aspect of 
the wake-field studies for the APT. Specifically, we apply the code 
ABCI \cite{ABCI} to compute longitudinal and transverse wakes in 
axisymmetric models of the APT 5-cell superconducting (SC) cavities 
using a varying number of bunches and looking for coherent 
wake-field effects. 

\section{Multiple-Bunch Effects in APT SC Cavities}

Wake potentials of a train of a few identical Gaussian bunches passing 
through a 5-cell APT SC cavity have been computed with the code ABCI 
\cite{ABCI}. Geometrical parameters of the APT cavities are given 
in \cite{FLK1}. The bunch rms length was chosen to be 4.5 mm in the 
$\beta$=0.82 section, and 3.5 mm for $\beta$=0.64. 
While these bunches have $\beta$=1, their separation is set to 
s=0.85657 m, which gives the proper bunch repetition frequency 
$f_b$=350 MHz. 

We study the loss factor for the 5-cell APT SC cavities 
as a function of the number of bunches $N_b$ in the bunch train. 
We expect that the loss factor per bunch tends to a constant for 
incoherent wakes, but it should increase linearly when wakes are 
coherent. The coherent effects occur if higher-mode resonances 
are close to multiples of $f_b$.
Correspondingly, the loss factor for the total train is proportional
to $N_b$ if there are no coherent effects, or increases faster,
up to $N^2_b$, otherwise. 

The results for the transverse loss factor $k_{tr}$ per bunch are 
shown in Fig.~1, both for $\beta$=0.64 and $\beta$=0.82. 
As one can see from Fig.~1, $k_{tr}$ reaches its asymptotic already 
for $N_b$ between 5 and 10 in the case of $\beta$=0.82. This 
asymptotic value, in fact, is lower than $k_{tr}$ for a single bunch. 
For $\beta$=0.64, however, we observe an almost linear growth up to 
$N_b$ about 20, and only after that the behavior changes and the 
transverse loss factor per bunch saturates. 
Therefore, in the $\beta$=0.64 cavity higher-order 
dipole resonances are closer to multiples of $f_b$ than those for 
$\beta$=0.82. 
For comparison, the total longitudinal loss factor for both cavities 
depends quadratically on $N_b$, while the loss factor per bunch 
increases linearly as $N_b$ increases. This is, of course, due to 
the fundamental accelerating mode of the cavity, whose frequency is 
700 MHz. 

\begin{figure}[htb]
\centerline{\epsfig{figure=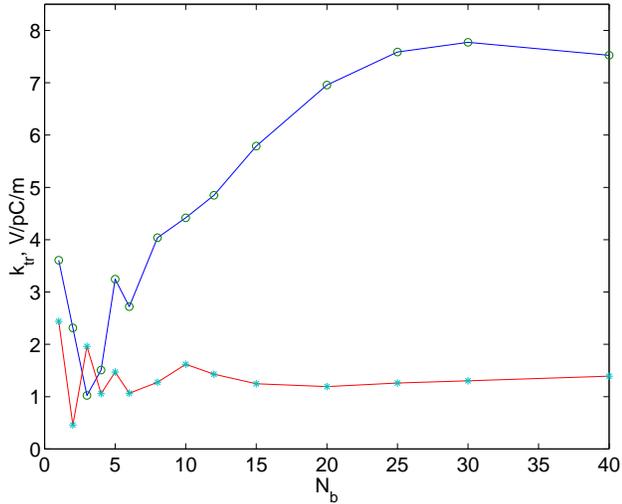,width=82.5mm}}
\caption{Transverse loss factor per bunch for 5-cell APT SC cavities 
versus the number of bunches: circles for $\beta=0.64$, 
stars for $\beta=0.82$.}
\end{figure}

The wake potentials for a bunch train with a monopole (on-axis beam) 
and dipole (off-axis beam) excitation look quite differently.
There is a strong coherent build-up of the amplitude of the 
longitudinal wake as long as bunches travel through the cavity. 
The bunches in the train interact with each other through the 
excitation of the cavity fundamental mode.
On the contrary, no apparent increase is observed for the transverse 
wake potential; wakes left by individual bunches are incoherent in 
this case. Therefore, one can use a maximal value of the transverse 
wake from these simulations as a reasonable estimate of that for a 
very large number of bunches, cf.\ Fig.~2. 
The maximum wakes from Fig.~2 allow to estimate the strength 
of beam-induced deflecting fields in the cavities for use in 
beam-dynamics simulations. 

\begin{figure}[htb]
\centerline{\epsfig{figure=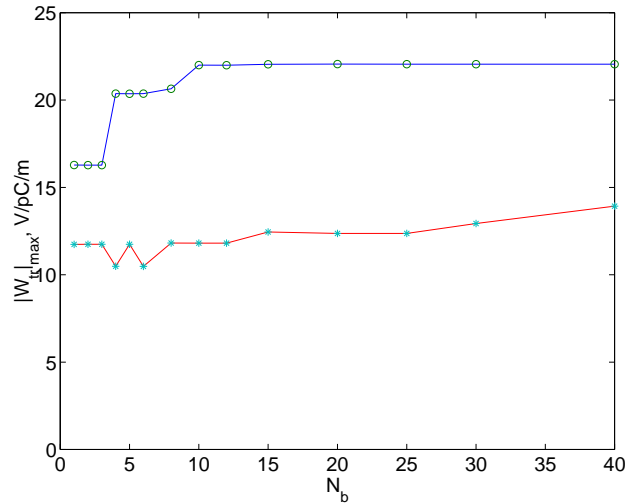,width=82.5mm}}
\caption{Maximum transverse wake potential for 5-cell APT SC 
cavities versus the number of bunches: circles for $\beta=0.64$, 
stars for $\beta=0.82$.}  
\end{figure}

To identify the frequency range where a higher dipole resonance in 
the APT SC $\beta$=0.64 5-cell cavity has its frequency close to the 
multiple of the bunch frequency $f_b$=350 MHz, we plot in Fig.~3 the 
power spectrum of the wake potential produced by a 30-bunch train in 
the cavity. One can see in Fig.~3 a regular structure of peaks 
at multiples of $f_b$, as well as a peak near 950 MHz, 
which corresponds to the band of the TM110 dipole mode  \cite{FLK2}. 
Comparison of the wake power spectra for different $N_b$
shows that the magnitude of this last peak decreases quickly as one 
goes to longer and longer bunch trains, since there is a smaller and 
smaller excitation at this frequency. However, 
it is the strong peak near 1750 MHz --- the multiple of the bunch 
frequency --- that produces a coherent increase of the dipole loss 
factor. Fortunately, its resonance frequency is close to the 
cutoff frequency of the pipe, which means this resonance can be 
effectively damped by HOM power couplers. 
Nevertheless, a more detailed analysis of this frequency range with 
frequency-domain codes is required to identify the corresponding 
eigenmode(s), and take its (their) properties into account in 
designing HOM couplers.

\begin{figure}[htb]
\centerline{\epsfig{figure=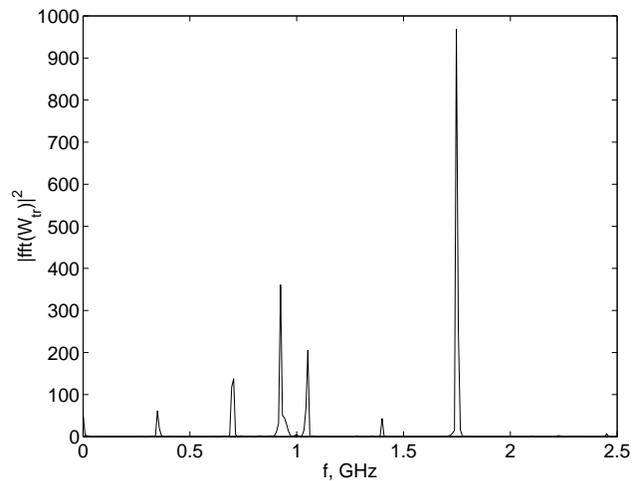,width=82.5mm}}
\caption{Power spectrum of the transverse wake potential for 30 
bunches in the 5-cell $\beta=0.64$ cavity.}  
\end{figure}

As the number of bunches in the train increases, its frequency 
spectrum is getting more and more concentrated near the multiples 
of the bunch repetition frequency. Stronger peaks in the wake power
spectrum for a relatively long bunch train indicate the frequency 
regions where the cavity resonances are close to multiples of $f_b$.
We show in Figs.~3-6 the power spectra of both the transverse 
and longitudinal wake potentials for 30-bunch trains. The wake 
potentials have been calculated for 30 m after the leading 
bunch in all cases, they include about 60,000 points, and 
their Fourier transforms have been performed with $N=2^{16}=64K$. 
A logarithmic scale is used for the longitudinal spectra, otherwise 
the pictures would be dominated completely by the cavity fundamental 
mode at 700 MHz. 

\begin{figure}[htb]
\centerline{\epsfig{figure=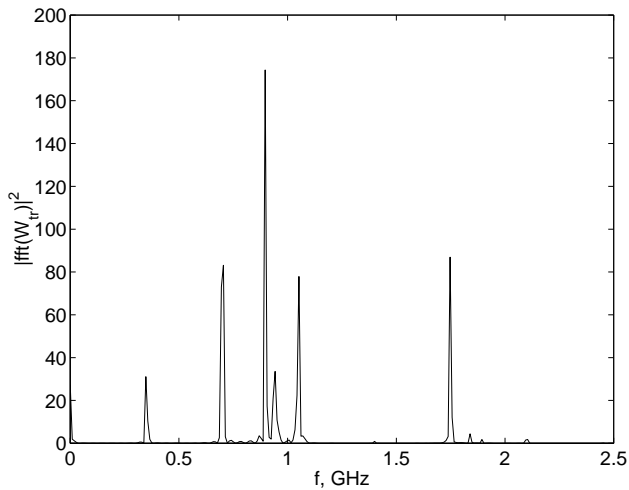,width=82.5mm}}
\caption{Power spectrum of the transverse wake potential for 30 
bunches in the 5-cell $\beta=0.82$ cavity.}
\end{figure}

\begin{figure}[htb]
\centerline{\epsfig{figure=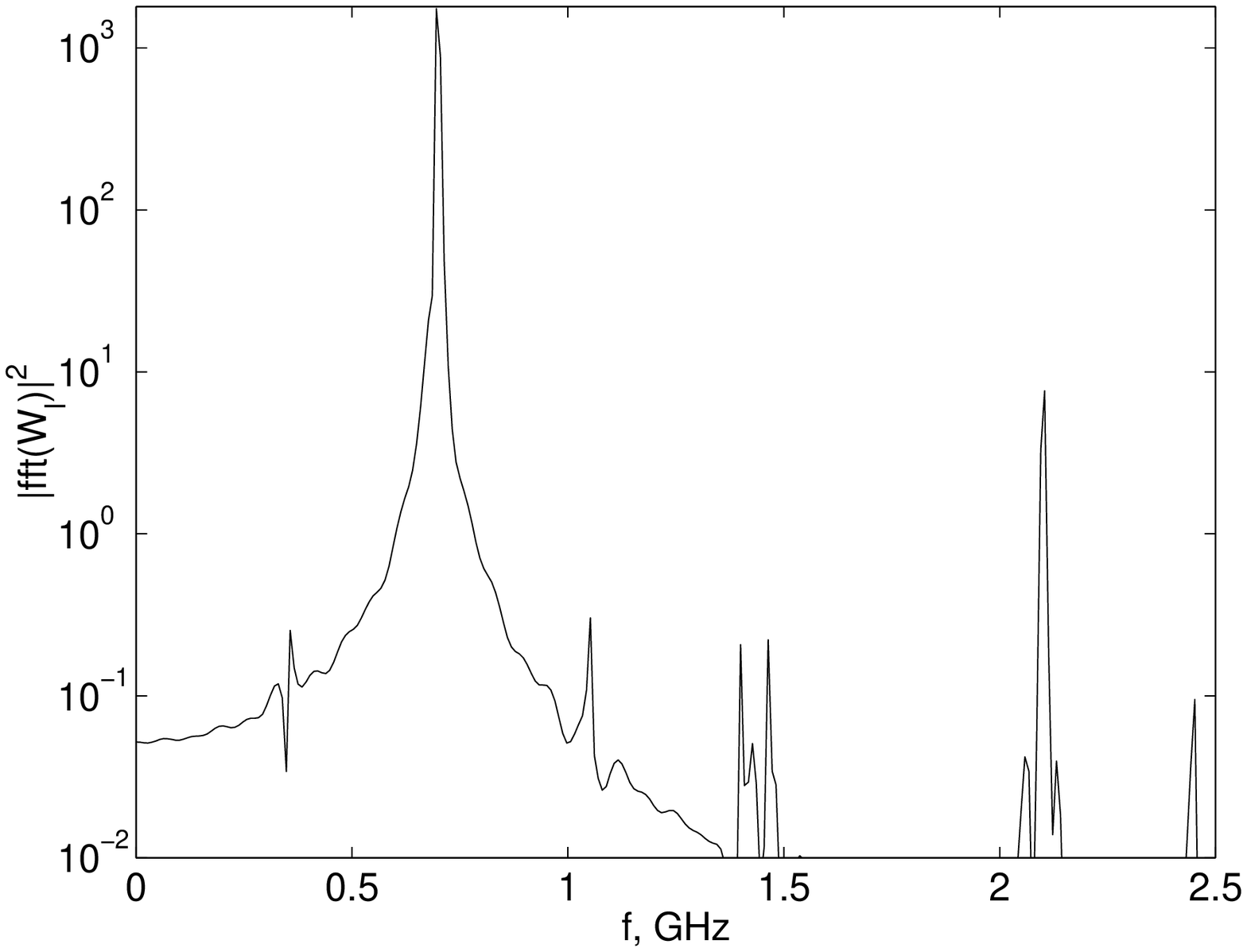,width=82.5mm}}
\caption{Power spectrum of the longitudinal wake potential for 30 
bunches in the 5-cell $\beta=0.64$ cavity.}
\end{figure}

\begin{figure}[htb]
\centerline{\epsfig{figure=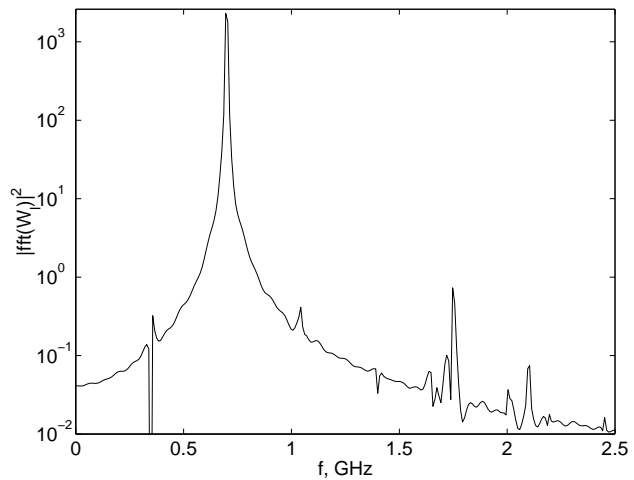,width=82.5mm}}
\caption{Power spectrum of the longitudinal wake potential for 30 
bunches in the 5-cell $\beta=0.82$ cavity.}
\end{figure}

Comparing relative peak heights in the frequency spectra shows 
where higher-order modes are close to multiples of the bunch frequency. 
Clearly, the potentially dangerous regions for the 5-cell $\beta$=0.64 
APT SC cavities are: around 1750 MHz and 1050 MHz with respect to the 
dipole modes; and near 2100 MHz for the monopole ones (of course, apart 
from 700 MHz). Since 2100 MHz is above the beam-pipe cutoff, one should
expect a trapped monopole mode near this frequency.
For 5-cell $\beta$=0.82 APT SC cavities these regions are: 
around 1750, 700, and 1050 MHz for the transverse modes (but all 
those contributions are relatively weak) and near 1750 and 1050 MHz 
for the longitudinal ones. Also, some additional attention is probably 
required to the transverse modes near 950 MHz for $\beta$=0.64 and in 
the range 900--950 MHz for the $\beta$=0.82 cavities. While these 
frequencies are not close to a multiple of $f_b$, the corresponding 
dipole resonances are strong enough that their effects are observed 
even for rather long bunch trains.

\section{Conclusions}

%Old abstr: A new method to study effects of multiple 
%non-ultrarelativistic 
%bunches passing a cavity is developed. It is based on time-domain 
%simulations for ultrarelativistic beams, and uses the corresponding 
%standard codes. This approach is then applied to the APT 
%superconducting cavities to identify potentially dangerous higher-
%order modes. 

A new approach to study higher-order mode effects in cavities for 
non-ultrarelativistic ($\beta < 1$) CW or long-pulse beams 
is proposed. It utilizes time-domain simulations using bunch trains 
which have $\beta$=1, but a correct bunch repetition frequency $f_b$. 
As the number of bunches $N_b$ increases, the details of the beam 
frequency spectrum, which are dependent both on $\beta$ and 
$N_b$, become unessential since the cavity is excited mostly at 
multiples of $f_b$. The approach allows applying standard time-domain 
codes, for example, \cite{ABCI}. 

Using this method we have found a few potentially dangerous frequency 
ranges of higher-order modes for the APT superconducting cavities. 
More details can be found in \cite{SK2}. 
A further analysis with frequency-domain codes is required to identify 
the modes in these frequency ranges, and to take their properties into 
account in designing HOM couplers.

The author would like to thank Frank Krawczyk and Thomas Wangler for 
useful discussions.


\begin{thebibliography}{9}
\bibitem{APT}
 G.P. Lawrence and T.P. Wangler, in Proceed. PAC97, Vancouver, BC,
1997; also in LA-UR-97-2582, Los Alamos, 1997.
\bibitem{ABCI}
 Y.H. Chin, Report LBL-35258, Berkeley, 1994.
\bibitem{MAFIA}
 MAFIA Release 4.00, CST, Darmstadt, 1997.
\bibitem{SK}
 S.S. Kurennoy, "Cavity Loss Factors for Non-Relativistic Beams", 
 Report LA-CP-98-55, Los Alamos, 1998; also in these proceedings.
\bibitem{FLK1}
 F.L. Krawczyk, et al., in Proceed. PAC97, Vancouver, BC,
1997; also LA-UR-97-1700, Los Alamos, 1997.
\bibitem{FLK2}
 F.L. Krawczyk, in Proceed. PAC97, Vancouver, BC,
1997; also LA-UR-97-1710, Los Alamos, 1997.
\bibitem{SK2}
 S.S. Kurennoy, "Multiple-Bunch Effects in APT SC Cavities", 
 Report LA-CP-98-151, Los Alamos, 1998.
\end{thebibliography}
\end{document}